\documentclass[10pt]{iopart}
\usepackage{iopams}
\usepackage{graphicx}
\usepackage[numbers,square,sort&compress]{natbib}
\usepackage[pdftex,colorlinks]{hyperref}
\usepackage{color}
\usepackage[shortcuts]{extdash}
\usepackage{upgreek}
\usepackage{bm}

\definecolor{darkblue}{rgb}{0,0,.6}
\hypersetup{colorlinks=true, breaklinks=true, citecolor=darkblue, linkcolor=darkblue, menucolor=darkblue, urlcolor=darkblue} 

\newcommand{\RuCl}{$\alpha$\=/RuCl$_3$}
\newcommand{\NaIrO}{Na$_2$IrO$_3$}
\newcommand{\jeff}{$j_{\rm eff}=1/2$}
\newcommand{\JEFF}{$j_{\rm eff}=3/2$}
\newcommand{\degC}{$^{\circ}\textrm{C}$}
\newcommand{\degr}{$^{\circ}$}
\newcommand{\tg}{$t_{\rm 2g}$}
\newcommand{\eg}{$e_{\rm g}$}
\newcommand{\Dtrig}{$D_{\rm trig}$}
\newcommand{\absDtrig}{$\left| D_{\rm trig} \right|$}
\newcommand{\Etal}{\textit{et al}}

\begin{document}

\title[RIXS study of $\alpha$-RuCl$_3$]{Resonant inelastic x-ray scattering study of $\alpha$-RuCl$_3$: a progress report}

\author{Blair W Lebert$^1$, Subin Kim$^1$, Valentina Bisogni$^2$, Ignace Jarrige$^2$, Andi M Barbour$^2$, Young-June Kim$^1$}

\address{$^1$ Department of Physics, University of Toronto, Toronto, Ontario, Canada M5S 1A7}
\address{$^2$ National Synchrotron Light Source II, Brookhaven National Laboratory, Upton, NY, 11973, USA}
\ead{\mailto{yjkim@physics.utoronto.ca}}
\vspace{10pt}
\begin{indented}
\item[]July 11th, 2017
\end{indented}

\begin{abstract}
Ru M$_3$-edge resonant inelastic x-ray scattering (RIXS) measurements of \RuCl{} with 27~meV resolution reveals a spin-orbit exciton without noticeable splitting. We extract values for the spin-orbit coupling constant ($\lambda=154\pm2$~meV) and trigonal distortion field energy ($\left|\Delta\right|<65$~meV) which support the \jeff{} nature of \RuCl{}. We demonstrate the feasibility of M-edge RIXS for $4d$ systems, which allows ultra high-resolution RIXS of $4d$ systems until instrumentation for L-edge RIXS improves.
\end{abstract}

\vspace{2pc}

\submitto{\JPCM}

\maketitle
\ioptwocol

\section{Introduction}\label{sec:introduction}

Study of Kitaev materials has blossomed in the last decade as a promising direction to find quantum spin liquids \cite{Kitaev2006}. The Kitaev honeycomb model is an exactly solvable model with quantum spin liquid (QSL) ground states \cite{Kitaev2006,Trebst2017,Winter2017,Hermanns2018}. One of the key developments in this field was the proposal by Jackeli and Khaliullin that the crucial ingredient of the Kitaev model, the bond-dependent Kitaev interaction, can be realized in a magnetic insulator with a strong spin-orbit coupling (SOC), such as the iridates where Ir$^{4+}$ ions take on a low-spin $d^5$ state due to large octahedral crystal field splitting \cite{Jackeli2009}. In particular, Chaloupka, Jackeli, and Khaliullin pointed out that a honeycomb lattice formed by edge-shared IrO$_6$ octahedra could be a realization of Kitaev's honeycomb model \cite{Chaloupka2010}. The successful synthesis of a honeycomb iridate compound \NaIrO{} \cite{Singh2010,Singh2012} and other iridate materials based on a honeycomb lattice or its three-dimensional variants opened a new genre of Kitaev materials research \cite{Modic2014,Takayama2015}, which has become an important branch of QSL research. For a comprehensive review on Kitaev materials, see Refs.~\cite{Winter2017,Takagi2019}.

One of the most promising Kitaev materials is \RuCl{}, which has been drawing much attention ever since it was first suggested as a candidate material for Kitaev QSL by Plumb \Etal{} \cite{Plumb2014}. Subsequent observation of a broad excitation continuum by inelastic neutron scattering \cite{Banerjee2017,Do2017} and Raman scattering \cite{Sandilands2015} suggests fractionalization of spin excitations expected for a QSL. Recently, the observation of quantized thermal Hall conductivity in the field-induced state of \RuCl{} has caused much excitement in the field of QSLs \cite{Kasahara2018}. 

The fact that \RuCl{} has emerged as a prime candidate for Kitaev QSL is somewhat surprising since its SOC is smaller than that of the iridates. However, one should realize that what is important is not necessarily the bare value of SOC, but the size of SOC in comparison to a non-cubic (i.e. trigonal or tetragonal) crystal field energy scale caused by the small distortions of RuCl$_6$ or IrO$_6$ octahedra. A large enough trigonal and/or tetragonal distortion could quench orbital angular momentum and spoil the Kitaev interaction between the \jeff{} pseudospins. Earlier structural report for \RuCl{} indicated that the RuCl$_6$ octahedron is free of trigonal distortion \cite{Stroganov1957}, which was one of the reasons behind the original proposal by Plumb \Etal{} \cite{Plumb2014}. However, recent powder neutron diffraction \cite{Banerjee2016} and single crystal x-ray diffraction \cite{Cao2016} studies revealed that the crystal structure is quite complicated due to stacking disorder, and most importantly, there exists small trigonal distortion. Therefore, it is in fact very important to re-examine whether the \jeff{} description survives this trigonal crystal field effect, which is a necessary condition for realizing Kitaev physics in \RuCl{}.

Resonant inelastic x-ray scattering (RIXS) has emerged as an invaluable technique for measuring electronic and magnetic excitations. In particular, RIXS is an exquisite tool for studying transitions between crystal-field-split $d$~orbitals. The electronic transitions between the $d$~orbitals, or $dd$ transitions, are dipole-forbidden and therefore typically difficult to study with optical spectroscopy. However, RIXS is a second-order process in which two dipole-allowed transitions are combined to provide a highly sensitive probe of $dd$ transitions. For example, Gretarsson \Etal{} measured the splitting of $dd$ transitions in \NaIrO{} and showed that the splitting due to the trigonal crystal field (0.1~eV) is much smaller than the overall splitting between the \jeff{} and \JEFF{} states, confirming the \jeff{} nature of its magnetism \cite{Gretarsson2013}. 

In this paper, we report our RIXS investigation of the $dd$ transitions in \RuCl{}. Similar to the \NaIrO{} case, our RIXS results show that the trigonal splitting is too small to observe with the current instrumental resolution. Our results thus suggest that the trigonal splitting is much smaller than the SOC energy scale, confirming the \jeff{} nature of this compound. Our result also indicates that the $g$ factor in \RuCl{} should be close to isotropic limit $g_\perp/g_\parallel\approx1$, in agreement with a recent x-ray absorption study \cite{Agrestini2017}.

We would like to also note that our measurements were carried out at the Ru M$_3$ edge, not at the L$_3$ edge which is more widely used for RIXS studies of cuprates and iridates. This study illustrates that using the M$_3$ edge instead of the L$_3$ edge is an option to be explored for the study of $4d$ elements, which allows researchers to use the existing soft x-ray RIXS beamlines instead of tender/intermediate energy range RIXS beamlines. We will discuss the advantages and disadvantages of using the M$_3$ edge instead of the L$_3$ edge. 

\section{Experimental}\label{sec:experimental}

Single crystals of \RuCl{} were prepared by vacuum sublimation of commerical RuCl$_3$ powder (Sigma-Aldrich, Ru content 45\% -- 55\%) in sealed quartz tubes, as in Ref. \cite{Plumb2014}. The three-zone tube furnace was cooled from 700\degC{} to 400\degC{} at 0.8\degC{}/hour while maintaining a 10\degC{} difference between the tube ends. The resulting platelike crystals have their crystallographic $c$ axis perpendicular to the surface and display hexagonal facets. Unless otherwise noted, we use the hexagonal notation corresponding to the the trigonal $P3_112$ space group from Ref. \cite{Stroganov1957} in this paper ($a=b=5.96$~\AA{}, $c=17.2$~\AA{}, $\alpha=\beta=90$\degr{}, $\gamma=120$\degr{}). The magnetic and thermodynamic properties of these single crystals have been well-characterized \cite{Sears2015}.

Single crystals were aligned and characterized using a four-circle x-ray diffractometer with a Mo-tube source. The sample we report on in this paper was relatively large, with a $\approx$5$\times$5~mm$^2$ area. Rocking curves taken at the (0,0,9) and (3,0,12) Bragg reflections showed a 1.5\degr{} mosaic. The crystal facets were confirmed to be high-symmetry hexagonal axes, which were used to align the sample during mounting. The sample was mounted on a copper holder using silver epoxy and cured for 3 hours at 80\degC{}. X-ray diffraction of the sample on the holder confirmed that the sample was aligned with $\langle$1\,0\,0$\rangle$ and $\langle$0\,0\,1$\rangle$ in the horizontal scattering plane.

We performed RIXS at the Ru M$_3$ edge which is at $\approx$462~eV, i.e. in the soft x-ray regime. Therefore, we used the SIX beamline \cite{Dvorak2016,Jarrige2018} at the NSLS-II. The VLS plane grating on the monochromator and spectrometer had line spacing of 500~mm$^{-1}$ and 1250~mm$^{-1}$ respectively, which coupled with a 20~$\upmu$m exit slit gave a combined resolution of 27~meV FWHM as measured on carbon tape. The x-ray beam size at the sample was 6 (H) $\times$ 13 (V) $\upmu$m$^2$. A $2\theta=90$\degr{} scattering angle and $\pi$-polarized incident x-rays were used to reduce the elastic line. The scattering angle gave a consant momentum transfer of $q=2k\sin{\theta}=0.33$~\AA$^{-1}$. All measurements reported here were performed at 20\degr{} grazing incidence giving q$_{\parallel}=0.14$~\AA$^{-1}$ in-plane. In terms of reciprocal lattice units (r.l.u) we measured in-plane at ${\bm q}=(0.13\,0)$, shown as a black diamond in the Brillouin zone inset of \fref{RIXS}. All measurements were performed at ambient temperature. The samples were cleaved with tape before putting the holder in the vacuum load-lock.

\section{Results}\label{sec:results}

\begin{figure}[t]
\includegraphics[width=\columnwidth]{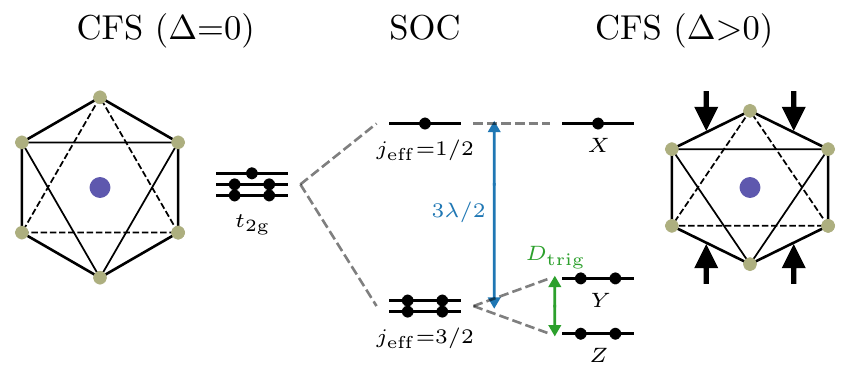}
\caption{\label{level} Schematic energy level diagram of Ru $4d$ orbitals in \RuCl{} under the influence of crystal field splitting (CFS) and spin-orbit coupling (SOC). On the left we show an ideal RuCl$_6$ octahedron, i.e. with no trigonal distortion ($\Delta=0$). Under this octahedral crystal field the degeneracy of the $4d$ energy levels is lifted and they split into \tg{} and \eg{} (not shown) manifolds. Turning on SOC splits the \tg{} manifold in \RuCl{} into a lower \JEFF{} level and higher \jeff{} level separated by $3\lambda/2$, where $\lambda$ is the SOC constant. A trigonal distortion field ($\Delta\neq0$) will lead to further splitting. Here we show the case of trigonal compression ($\Delta>0$) which is illustrated by the arrows compressing the RuCl$_6$ octahedron on the right. Under small trigonal compression the \JEFF{} levels split giving in the end three non-degenerate doublets. We have labeled these levels as $X$, $Y$, and $Z$ corresponding to $A$, $B$, and $C$ respectively in the notation of Chaloupka \Etal{} \cite{Chaloupka2016} to avoid confusion with our labeling of the RIXS peaks. The splitting of the lower two levels defines \Dtrig{} $= E_Y - E_Z$, which is therefore positive (negative) for trigonal compression (elongation).
}
\end{figure}

In \RuCl{}, the Ru$^{3+}$ ions are surrounded by a nearly ideal octahedral arrangement of Cl$^-$ ions (\fref{level}). These RuCl$_6$ octahedra share an edge to form a honeycomb net. The octahedral cubic crystal field, $10Dq$, splits the Ru $4d$ levels into lower \tg{} and upper \eg{} manifolds. It is believed that SOC further splits the \tg{} manifold into upper \jeff{} and lower \JEFF{} states. The $4d^5$ electronic configurations of the Ru ions should take on a low-spin state, filling up the \JEFF{} level completely leaving a half-filled \jeff{} level and leading to the so-called \jeff{} pseudospin. The RuCl$_6$ octahedra however have a slight trigonal distortion which might quench the oribital angular momentum and destroy the \jeff{} nature of \RuCl{}. 

X-ray absorption spectroscopy (XAS) is a useful tool to shed light on the influence of SOC on electronic structure. XAS measurements by Plumb \Etal{} \cite{Plumb2014} on \RuCl{} at the Ru L$_2$ ($2p_{1/2}$) and L$_3$ ($2p_{3/2}$) edges observed a difference in their lineshape and found an anomalously large L$_3$/L$_2$ intensity ratio, the so-called branching ratio (BR). Assuming nonnegligible SOC, atomic dipole transitions must follow the $J$ selection rules. Therefore at the L$_2$ edge, $2p_{1/2}\rightarrow4d_{3/2}$ is allowed and $2p_{1/2}\rightarrow4d_{5/2}$ is forbidden, while at the L$_3$ edge, $2p_{3/2}\rightarrow4d_{3/2}$ and $2p_{3/2}\rightarrow4d_{5/2}$ are both allowed. This is manifested in the XAS measurements of Plumb \Etal{}, where they find the L$_2$ absorption edge has a single peak, compared to double peaks at the L$_3$ edge. Of these two peaks, the lower (higher) energy one is due to \tg{} (\eg{}) states and correspondingly has a lower (higher) intensity because of its one (four) hole(s). The absence of a \tg{} peak at the L$_2$ edge is because it acquires a $J=5/2$ character due to SOC. The BR is one way of expressing this effect of SOC, where a BR $>2$ is expected with SOC effects due to a reduction of dipole-allowed transitions for the L$_2$ edge as explained above. BR $=2$ is typical without SOC since the L$_3$ edge has twice as many electrons available as the L$_2$ edge: $2p_{3/2}$ is a quartet and $2p_{1/2}$ is a doublet. Plumb \Etal{} reported a BR $=3\pm0.5$ for \RuCl{}.

\begin{figure*}[t]
\includegraphics[width=\textwidth]{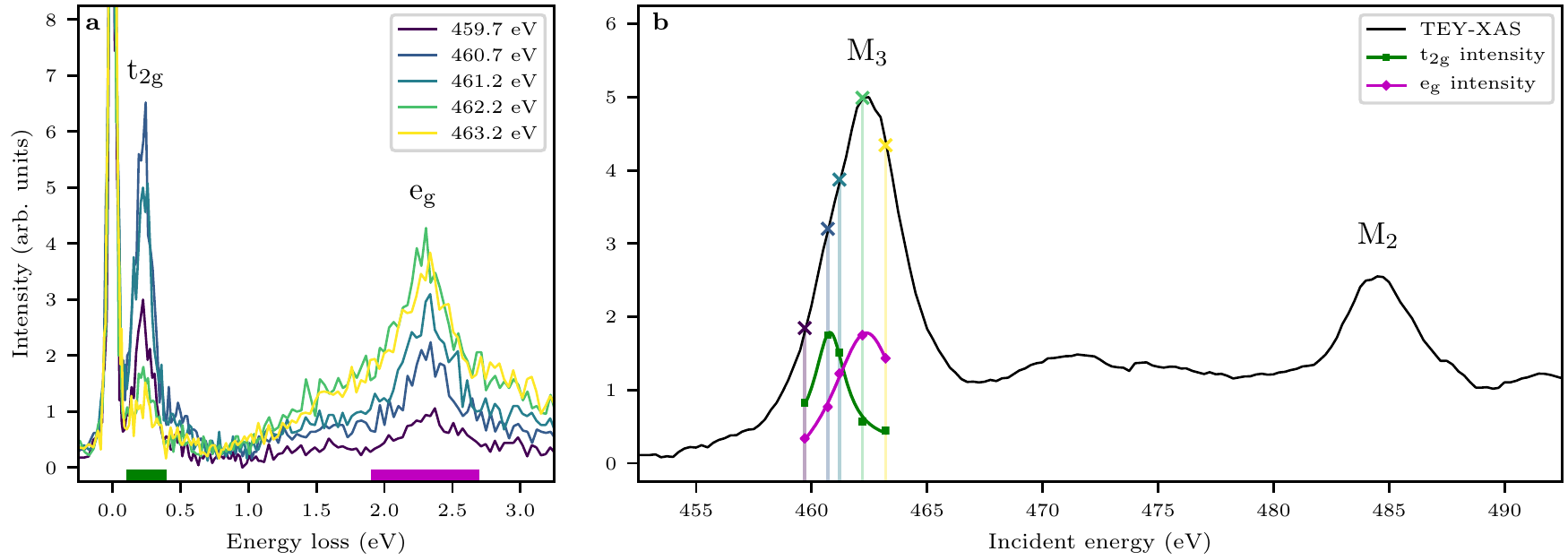}
\caption{\label{Edep} \textbf{(a)} Ru M$_3$-edge RIXS spectra on \RuCl{} as a function incident energy, $E_i$. The lower (higher) energy features resonate at a lower (higher) energy and correspond to \tg{} (\eg{}) excitations. The green and magenta bars represent the energy range integrated over to extract the intensity for \tg{} and \eg{} respectively. \textbf{(b)} Total electron yield x-ray absorption spectroscopy (TEY-XAS) measurement on \RuCl{}. The integrated intensities of the \tg{} and \eg{} RIXS regions are plotted as a function of $E_i$ in green and magenta respectively.
}
\end{figure*}

We performed XAS at the Ru M$_2$ ($3p_{1/2}$) and M$_3$ ($3p_{3/2}$) edge in total electron yield (TEY) mode, shown as a black line in \fref{Edep}(b). The first noticeable difference is that the \tg{}-\eg{} double peak structure is not visible at the M$_3$ edge. The M edges correspond to shallower core holes than the L edges, i.e. lower binding energy, and therefore have shorter lifetimes. The decreased lifetime leads to broadening in the energy domain which make it appear as a single peak. Nonetheless, the peak appears asymmetric and has a slight shoulder near 460.5~eV, which with the main peak at 462.5~eV would give an estimate of the splitting between \tg{} and \eg{} of $\approx2.0$~eV with a large error bar. In comparison, a splitting of $\approx2.3$~eV was observed in the L-edge XAS data \cite{Plumb2014}. We also roughly estimated the BR by subtracting a linear background at each peak and integrating the intensity. We found BR $=3.6\pm0.8$ which is consisent with the BR from L-edge XAS \cite{Plumb2014}.

We first studied the resonant behavior of RIXS by scanning the incident energy $E_i$ as shown in \fref{Edep}(a). We measured RIXS spectra at five different energies, which are also shown relative to the TEY-XAS signal as vertical lines with the same color in \fref{Edep}(b). The Ru M$_3$-edge RIXS directly probes the crystal-field split Ru $4d$ levels, the so-called $dd$ excitations. The two dipole transitions, $3p_{3/2} \rightarrow 4d$ followed by $4d \rightarrow 3p_{3/2}$, allow dipole-forbidden transitions from occupied \tg{} states to empty \tg{} and \eg{} states. The resonant enhancement is indeed strong, over an order of magnitude, and we find two different resonance regimes. The excitations to unoccupied \tg{} states correspond to the spectral weight below $<1$~eV while excitations to \eg{} states occur above this energy. We quantitatively studied this resonant behavior by integrating the spectral weight in each region. The integration ranges are shown as solid bars at the bottom of \fref{Edep}(a): the \tg{} region in green from 0.1 to 0.4~eV and the \eg{} region in magenta from 1.9 to 2.7~eV. The integrated intensities are plotted in arbitrary units in \fref{Edep}(b) as green squares for \tg{} and magenta diamonds for \eg{} (the five points are fit with a Lorentzian function and constant background as shown by the corresponding solid lines). The \tg{} and \eg{} states resonate at approximately 460.8~eV and 462.4~eV respectively. The $\approx1.6$~eV difference between these energies corresponds roughly to the $10Dq$ octahedral crystal-field energy, estimated to be $\approx2.0$~eV earlier.

\begin{figure}[t]
\includegraphics[width=\columnwidth]{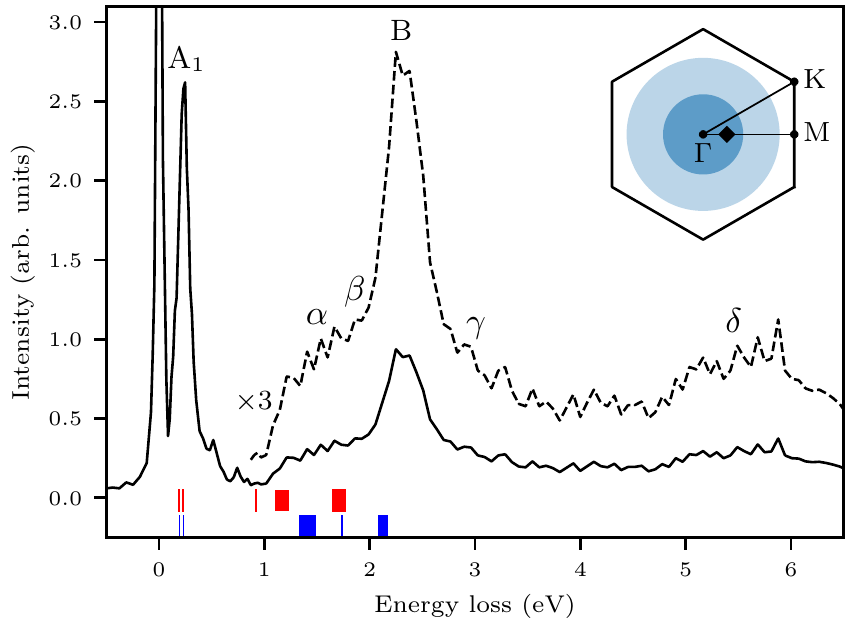}
\caption{\label{RIXS} Ru M$_3$-edge RIXS ($E_i=461$~eV) spectrum of \RuCl{} single crystals measured at 300~K. A $2\theta=90$\degr{} scattering angle was used and the incident beam was in 20\degr{} grazing incidence with the sample, giving a measurement at ${\bm q}=(0.13\,0)$ r.l.u. (black diamond in Brillouin zone inset). The features are labeled using the notation of Sandilands \Etal{} \cite{Sandilands2016}. Predicted $dd$ excitation energies from quantum chemistry calculations \cite{Yadav2016} are shown for the $P3_112$ (red) \cite{Stroganov1957} and $C2/m$ (blue) \cite{Cao2016} structures. The inset shows  the Brillouin zone of \RuCl{} in hexagonal notation. The dark (light) blue represent areas which can be probed by Ru M$_3$-edge RIXS with $2\theta=90$\degr{} ($2\theta=150$\degr{}).
}
\end{figure}

\begin{figure*}[t]
\includegraphics[width=\textwidth]{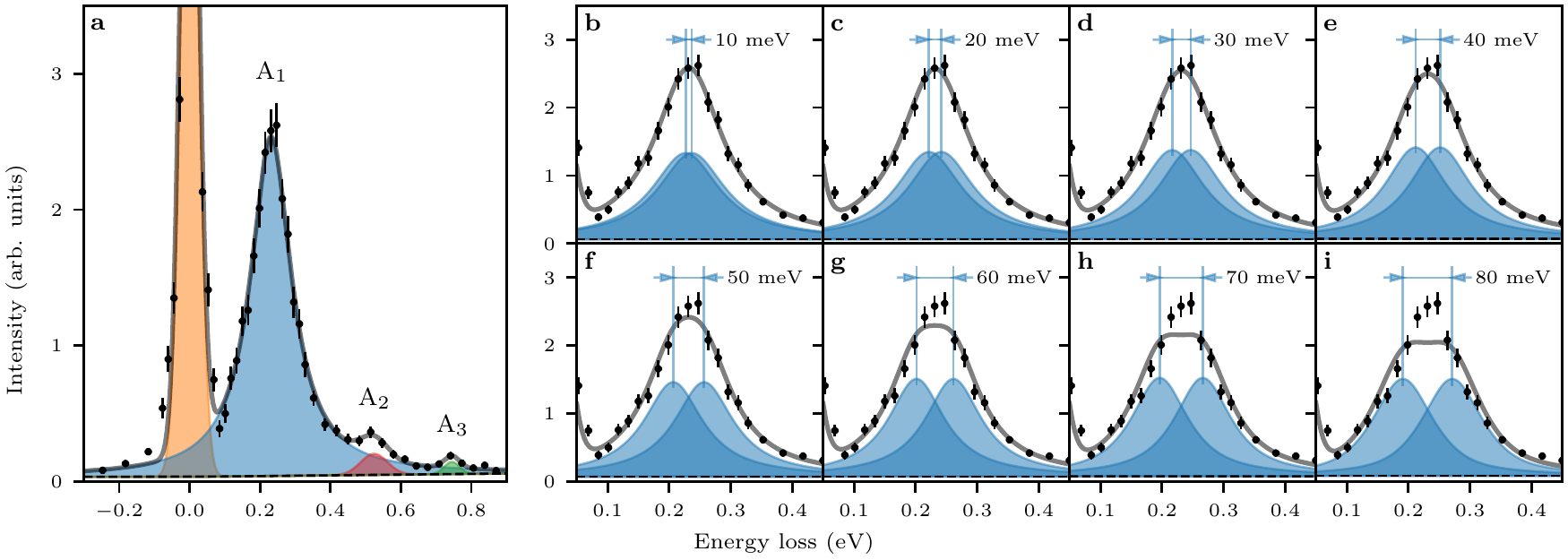}
\caption{\label{fit} \textbf{(a)} Low-energy region of the RIXS spectrum from \fref{RIXS}. The elastic (orange), A$_2$ (red), and A$_3$ (green) peaks are fit with Gaussian functions while the A$_1$ peak is fit with a Lorentzian function. A constant background and linear background only in the energy loss region are also included in the fit (dashed black line). The energies extracted from the fit are 233~meV, 529~meV, and 743~meV for A$_1$, A$_2$, and A$_3$. \textbf{(b--i)} Fit of the A$_1$ peak using two peaks separated by a fixed \absDtrig{} value. The width and intensity of the peaks varies for different \absDtrig{} values, but the two peaks are constrained to the same values for each individual fit. The peaks are pseudo-Voigt functions, where both peaks are fixed with the same Lorentzian/Gaussian ratio. The peaks are best fit with 100\% Lorentzian for 10--60~meV, 93\% for 70~meV, and 78\% for 80~meV.
}
\end{figure*}

To search for splitting due to trigonal distortions we focused on the \tg{} region. Therefore, we measured a high statistics spectrum with $E_i=461$~eV, i.e. at the resonance of the \tg{} states. In \fref{RIXS}, we show a high-resolution Ru M$_3$-edge RIXS spectrum on \RuCl{} which has many features visible in the energy loss region: three low-energy features A$_1$, A$_2$, and A$_3$ (zoom in \fref{fit}) at $231\pm3$~meV, $524\pm10$~meV, and $745\pm10$~meV; a charge gap around 1~eV; and, high-energy features $\alpha$, $\beta$, B, $\gamma$, and $\delta$ above 1~eV. The features have been labeled with the same notation as Sandilands \Etal{} \cite{Sandilands2016} with the addition of the B peak.

The energies of the $dd$ excitations are consistent with those found with optical spectroscopy \cite{Sandilands2016} and quantum chemistry calculations \cite{Yadav2016}. The quantum chemistry calculation results are indicated in \fref{RIXS}. Yadav and coworkers calculated $dd$ transition energies using multireference configuration-interaction (MRCI) calculations with SOC for two different structures: $P3_112$ \cite{Stroganov1957} shown in red and $C2/m$ \cite{Cao2016} shown in blue. Besides the B peak, the other high-energy RIXS features, $\alpha$, $\beta$, and $\gamma$, are broad and difficult to compare with calculations. Nonetheless, the $\alpha$, $\beta$, and B peaks correspond well with predicted energies for the $C2/m$ structure, although the RIXS measurements are at a slightly higher energy. On the other hand, the $P3_112$ structure predicts energies which are far too low compared to the RIXS data. Recent structural studies seem to agree on the fact that the room temperature structure is described by the $C2/m$ symmetry \cite{Cao2016,Johnson2015,Sears-thesis}. Our measurements performed at 300~K clearly agrees with the $C2/m$ structure prediction.

The low-energy features of the RIXS spectrum are shown in \fref{fit}(a). The A$_1$ peak is a $dd$ excitation from \JEFF{} to \jeff{}, and is therefore also known as a spin-orbit exciton \cite{Jungho2012}. Ignoring trigonal distortions, the A$_1$ peak position corresponds to the energy difference between these levels, $231\pm3$~meV. The splitting between these levels is equal to $3\lambda/2$ (\fref{level}) from which we find a SOC constant of $\lambda=154\pm2$~meV. There is no apparent energy splitting due to trigonal distortion --- the A$_1$ excitation is best fit with a single Lorentzian peak. The low energy region was fit with four peaks as shown in \fref{fit}(a). The elastic line (orange), A$_2$ (red), and A$_3$ (green) peaks were fit with Gaussian functions, while the A$_1$ peak (blue) was fit with a Lorentzian function. The fit also included a background function, shown as a dashed black line, which included a constant offset with a linear slope only in the energy loss region.

We also performed a series of fits of the A$_1$ feature using two peaks with fixed splitting, shown in \fref{fit}(b--i), to estimate an upper bound on the trigonal splitting \absDtrig{} (\fref{level}). Each panel corresponds to a fit using a fixed \absDtrig{} ranging from 10 to 80~meV. The relative energy position of the peaks is fixed by this splitting, however they are free to move in energy together during the fit. Both peaks are described by a pseudo-Voigt function with the same intensity, width, and Lorentzian/Gaussian ratio, however these values are fit for each of the individual \absDtrig{} values. We notice the fit diverging from our data at \absDtrig{} $<40$~meV which we define as our upper bound. 

The energy difference between the trigonally split \JEFF{} levels is given by \absDtrig{}, however this is not equivalent to the trigonal field energy $\Delta$. Furthermore, the sign of both of these values depends on whether it is trigonal compression ($\Delta>0$, \Dtrig{} $>0$) or elongation ($\Delta<0$, \Dtrig{} $<0$). In \fref{level} we show the case of trigonal compression where the energy level of $Y$ is above $Z$, while in the case of trigonal elongation the opposite would be true. Our current measurements cannot differentiate between compression or elongation, therefore we have to estimate $\Delta$ separately for each case. The splitting between the levels is given by \Dtrig{}$/\lambda=\frac{1}{4}[\sqrt{8+(1+\delta)^2} - 3 + \delta]$, where $\delta=2\Delta/\lambda$ \cite{Chaloupka2016}. From this we find for compression $\left|\Delta\right|<55$~meV and for elongation $\left|\Delta\right|<65$~meV

\section{Discussion}\label{sec:discussion}

We have confirmed that the trigonal splitting in \RuCl{} is small enough, such that the energy hierarchy needed for a \jeff{} ground state is satisfied: $\left|\Delta\right|$ ($<$0.065~eV) $\ll$ $\lambda$ (0.154~eV) $\ll$ $10Dq$ ($\approx$2~eV). Our results are consistent with the MRCI calculations which predict a splitting of \absDtrig{} $=39$~meV for the $C2/m$ structure. Ru L-edge XAS reports a vanishingly small linear dichroism effect corresponding to \Dtrig{} $=-12 \pm 10$~meV \cite{Agrestini2017}. It is interesting to note this report of trigonal elongation is at odds with structural studies \cite{Cao2016,Johnson2015,Sears-thesis} which show trigonal compression. Regardless of the sign of \Dtrig{}, the small magnitude found in this study and by Agrestini \Etal{} indicates the $g$ factor should be nearly isotropic.

The limit \absDtrig{} $<40$~meV is likely an overestimation since our energy resolution was 27~meV and we  observed no splitting of the spin-orbit exciton. The use of equal intensity peaks for our fitting in \fref{fit}(b--i) is an approximation, the peaks should actually have different relative intensities depending on the RIXS matrix elements. Therefore, future high-resolution Ru M$_3$-edge RIXS studies on \RuCl{} can use different polarizations and incident angles to determine more accurately the magnitude and perhaps even the sign of \Dtrig{} \cite{Chaloupka2016, Jungho2014}.

Optical spectroscopy observes low-energy features at similar energies as our RIXS results: A$_1\approx270$~meV, A$_2\approx530$~meV, and A$_3\approx740$~meV at room temperature \cite{Sandilands2016, Borgwardt2019}. Additionally, an extremely sharp peak was observed with Raman spectroscopy at A$_0\approx145$~meV ($\lambda\approx96$~meV) \cite{Sandilands2016}. The A$_0$ peak was initially attributed as the spin-orbit exciton, however it is at lower energy than our RIXS results and MRCI calculations \cite{Yadav2016}. The 2~meV width of the A$_0$ Raman mode was a surprise since coupling with phonons is expected to broaden the spin-orbit exciton peak \cite{Plotnikova2016}. Furthermore, a recent Raman spectroscopy study on \RuCl{} finds that the A$_0$ peak vanishes with increasing temperatures, which is unexpected behavior for a spin-orbit exciton \cite{Borgwardt2019}. This study also reports on a higher energy Raman mode at $\approx235$~meV which agrees very well with our RIXS results. Inelastic neutron scattering has observed hints of an excitation around this energy region, albeit with poor statistics \cite{Banerjee2016}. Curiously, the SOC constant we report here, $\lambda=154\pm2$~meV, agrees almost exactly with the tabulated Ru$^{3+}$ free ion value ($\lambda=155$~meV) \cite{Porterfield2013}.

Previous studies incorrectly assigned the A$_1$--A$_3$ peaks as transitions to SOC-split \eg{} states \cite{Plumb2014,Sandilands2016,Reschke2017}. The resonant behavior we observe, coupled with MRCI calculations \cite{Yadav2016}, demonstrates that their energy is too low to correspond to \eg{} excitations. The A$_1$ peak observed with optical spectroscopy is the spin-orbit exciton but since it is dipole-forbidden it is a phonon-assited transition. Indeed, the optical peak is shifted 30--40~meV higher in energy, which is consistent with the energy of optical phonons measured in \RuCl{} \cite{Sandilands2015}. The temperature dependence of all three peaks was studied by Sandilands \Etal{} \cite{Sandilands2016}, integrating the spectral weight from 0.1--0.87~eV, and was consistent with a phonon-assisted mechanism. However, the A$_1$ peak is dominant in this region and Borgwardt \Etal{} \cite{Borgwardt2019} found that the A$_2$ and A$_3$ peak do not have a strong temperature dependence, i.e. A$_2$ and A$_3$ are not phonon-assisted transitions. This is shown clearly by comparing their energies ($\approx530$~meV and $\approx740$~meV) to our RIXS energies (523~meV and 745~meV), which are nearly identical. 

Borgwardt \Etal{} \cite{Borgwardt2019} interpreted these peaks as multiparticle excitations, i.e. double and triple spin-orbit excitons, but this seems unlikely. For example, the energy difference between multiples of a single excitation and a multiple excitation should correspond to a phonon energy, but ${\rm A}_2 - 2\cdot{\rm A}_1\approx60$~meV which is higher than any observed phonon in \RuCl{}. As well, we observe the same peaks with RIXS as optical spectroscopy strongly indicating a common origin. Multiparticle excitations are a nonlinear effect and thus their observation usually requires high-intensity photon beams. This could be possible in RIXS or Raman spectroscopy, however the infrared absorption experiment used a low-intensity lamp \cite{Sandilands2016}. 

We are not completely certain of the true nature of these two peaks, however we believe that Ru $4d$-Cl $3p$ hybridization plays a key role. One picture could be Cl $3p$ $\rightarrow$ Ru $4d$ charge-transfer type excitations. For example, recent density functional theory calculations found two sharp peaks in the DOS with significant Ru-Cl hybridization at 590~meV and 730~meV above the Fermi level \cite{Agrestini2017}.

To our knowledge, this is the first report of M-edge RIXS on a $4d$ transition metal system. Actually, there are very few reports of M-edge RIXS at all. The first M-edge RIXS experiment cleverly leveraged the low energy of the Cu M$_{2,3}$ edge to achieve improved resolution on existing instrumentation \cite{Kuiper1998} before the development of next-generation soft x-ray RIXS beamlines \cite{Dvorak2016,Jarrige2018,Lai2014,Brookes2018,I21,Veritas}. Studying $3d$ transition metal systems with M-edge RIXS is rare since the lifetime of the shallow core hole has a complex dependence on incident photon energy, the elastic line due to off-specular reflectivity is extremely strong and obscures low-energy features, and the inelastic cross-section is lower due to increased Auger emission \cite{Wray2015}.

These first two disadvantages are related to the $<120$~eV incident energy used for M-edge RIXS of $3d$ systems and not so important for $4d$ systems which have higher energy M edges. In our experiment, the elastic line intensity was three times the A$_1$ intensity, but we measured at 300~K and we found considerable quasi-elastic weight (50~meV FWHM vs. 27~meV FWHM resolution). These results are promising but further studies at higher resolutions and lower temperatures, as well as varying the scattering angle away from $2\theta=90$\degr{}, are important to determine if elastic intensity will be a limiting factor of M-edge RIXS in $4d$ systems. The inelastic cross-section is generally lower for M vs. L edges due to the decreased fluorescence yield. For example, in Ru the fluorescence yield is $2.3\cdot10^{-4}$ and $4.5\cdot10^{-2}$ for the M$_3$ and L$_3$ edge respectively \cite{Bambynek1972}, however our results show that the resonant enhancement is still sufficient to perform experiments. 

One disadvantage is that the lower energy of the M edge limits the area of the Brillouin zone which can be probed. The inset of \fref{RIXS} shows the area available with Ru M$_3$-edge RIXS in \RuCl{} as dark (light) blue for $2\theta=90$\degr{} ($2\theta=150$\degr{}), while Ru L$_3$-edge RIXS can probe a few Brillouin zones. We note that the accessible Brillouin zone in \RuCl{} at the M edge is still enough to search for the gapless Majorana fermions predicted by theory \cite{Halasz2016}. As well, in general the larger supercells of magnetically ordered materials will have a correspondingly smaller magnetic Brillouin zone which could possibly be probed completely at the M edge. However, the decreased SOC and lifetime of the $3p$ core hole in M-edge RIXS does make it less effective at measuring magnons \cite{Ament2011}.

Nonetheless, M-edge RIXS for $4d$ systems has the enormous advantage that many soft x-ray RIXS beamlines already exist or are under development which can provide high flux and sub-30~meV resolution \cite{Dvorak2016,Jarrige2018,Lai2014,Brookes2018,I21,Veritas}. For Ru L$_3$-edge RIXS (2840~eV) there is currently only one instrument in the world, IRIXS at P1/DESY, which is currently unavailable to general users. The latest results from IRIXS \cite{Suzuki2019,Gretarsson2019} have a resolution approximately a factor of 5 worse than the resolution routinely available for Ru M$_3$-edge RIXS. 

\section{Conclusion}\label{sec:conclusion}
We have performed Ru M$_3$-edge resonant inelastic x-ray scattering (RIXS) on \RuCl{}. We observe $dd$ excitations in agreement with optical spectroscopy and quantum chemistry calculations. Our observation of a spin-orbit exciton allows us to extract a very accurate value for the spin-orbit coupling constant $\lambda=154\pm2$~meV. The spin-orbit exciton shows no splitting due trigonal distortions and overall we find the energy hierarchy necessary for \jeff{} physics is satisfied in \RuCl{}. Our results resolve some previous misconceptions about the electronic structure of \RuCl{} and provide a springboard for future calculations and experiments to further elucidate its nature. Measurement of $4d$ systems with M-edge RIXS is a novel technique which we believe will be an important part of the x-ray spectroscopist's toolbox since it allows ultra high-resolution RIXS measurements here and now.

\ack
Work at the University of Toronto was supported by the Natural Science and Engineering Research Council
(NSERC) of Canada, Canadian Foundation for Innovation, and Ontario Innovation Trust. This research used the SIX beamline of the National Synchrotron Light Source II, a U.S. Department of Energy (DOE) Office of Science User Facility operated for the DOE Office of Science by Brookhaven National Laboratory under Contract No. DE-SC0012704. This work was performed in part at Aspen Center for Physics, which is supported by National Science Foundation grant PHY-1607611. BWL acknowledges the support from the University of Toronto Faculty of Arts and Sciences Postdoctoral Fellowship. 

\bibliographystyle{iopart-num}
\bibliography{main}

\providecommand{\newblock}{}
\begin{thebibliography}{10}
\expandafter\ifx\csname url\endcsname\relax
  \def\url#1{{\tt #1}}\fi
\expandafter\ifx\csname urlprefix\endcsname\relax\def\urlprefix{URL }\fi
\providecommand{\eprint}[2][]{\url{#2}}

\bibitem{Kitaev2006}
Kitaev A 2006 {\em Annals of Physics\/} {\bf 321} 2 -- 111 ISSN 0003-4916
  january Special Issue

\bibitem{Trebst2017}
Trebst S 2017 {\em arXiv preprint arXiv:1701.07056\/}

\bibitem{Winter2017}
Winter S~M, Tsirlin A~A, Daghofer M, van~den Brink J, Singh Y, Gegenwart P and
  Valenti R {2017} {\em {Journal of Physics - Condensed Matter}\/} {\bf {29}}
  ISSN {0953-8984}

\bibitem{Hermanns2018}
Hermanns M, Kimchi I and Knolle J 2018 {\em Annual Review of Condensed Matter
  Physics\/} {\bf 9} 17--33

\bibitem{Jackeli2009}
Jackeli G and Khaliullin G 2009 {\em Phys. Rev. Lett.\/} {\bf 102}(1) 017205

\bibitem{Chaloupka2010}
Chaloupka J, Jackeli G and Khaliullin G 2010 {\em Phys. Rev. Lett.\/} {\bf
  105}(2) 027204

\bibitem{Singh2010}
Singh Y and Gegenwart P 2010 {\em Phys. Rev. B\/} {\bf 82} 064412

\bibitem{Singh2012}
Singh Y, Manni S, Reuther J, Berlijn T, Thomale R, Ku W, Trebst S and Gegenwart
  P 2012 {\em Phys. Rev. Lett.\/} {\bf 108}(12) 127203

\bibitem{Modic2014}
{Modic} K~A, {Smidt} T~E, {Kimchi} I, {Breznay} N~P, {Biffin} A, {Choi} S,
  {Johnson} R~D, {Coldea} R, {Watkins-Curry} P, {McCandless} G~T, {Chan} J~Y,
  {Gandara} F, {Islam} Z, {Vishwanath} A, {Shekhter} A, {McDonald} R~D and
  {Analytis} J~G 2014 {\em Nature Communications\/} {\bf 5} 4203

\bibitem{Takayama2015}
Takayama T, Kato A, Dinnebier R, Nuss J, Kono H, Veiga L~S~I, Fabbris G, Haskel
  D and Takagi H 2015 {\em Phys. Rev. Lett.\/} {\bf 114}(7) 077202

\bibitem{Takagi2019}
Takagi H, Takayama T, Jackeli G, Khaliullin G and Nagler S~E 2019 {\em Nature
  Reviews Physics\/} {\bf 1} 264--280

\bibitem{Plumb2014}
Plumb K~W, Clancy J~P, Sandilands L~J, Shankar V~V, Hu Y~F, Burch K~S, Kee H~Y
  and Kim Y~J 2014 {\em Phys. Rev. B\/} {\bf 90}(4) 041112

\bibitem{Banerjee2017}
Banerjee A, Yan J, Knolle J, Bridges C~A, Stone M~B, Lumsden M~D, Mandrus D~G,
  Tennant D~A, Moessner R and Nagler S~E 2017 {\em Science\/} {\bf 356}
  1055--1059 ISSN 0036-8075

\bibitem{Do2017}
Do S~H, Park S~Y, Yoshitake J, Nasu J, Motome Y, Kwon Y~S, Adroja D~T, Voneshen
  D~J, Kim K, Jang T~H, Park J~H, Choi K~Y and Ji S 2017 {\em Nat. Phys.\/}
  {\bf 13} 1079--1084

\bibitem{Sandilands2015}
Sandilands L~J, Tian Y, Plumb K~W, Kim Y~J and Burch K~S 2015 {\em Phys. Rev.
  Lett.\/} {\bf 114}(14) 147201

\bibitem{Kasahara2018}
Kasahara Y, Ohnishi T, Mizukami Y, Tanaka O, Ma S, Sugii K, Kurita N, Tanaka H,
  Nasu J, Motome Y, Shibauchi T and Matsuda Y 2018 {\em Nature\/} {\bf 559}
  227--231 ISSN 1476-4687

\bibitem{Stroganov1957}
Stroganov E~V and Ovchinnikov K~V 1957 {\em Ser. Fiz. i Khim.\/} {\bf 12} 152

\bibitem{Banerjee2016}
{Banerjee} A, {Bridges} C~A, {Yan} J~Q, {Aczel} A~A, {Li} L, {Stone} M~B,
  {Granroth} G~E, {Lumsden} M~D, {Yiu} Y, {Knolle} J, {Bhattacharjee} S,
  {Kovrizhin} D~L, {Moessner} R, {Tennant} D~A, {Mandrus} D~G and {Nagler} S~E
  2016 {\em Nature Materials\/} {\bf 15} 733--740

\bibitem{Cao2016}
Cao H~B, Banerjee A, Yan J~Q, Bridges C~A, Lumsden M~D, Mandrus D~G, Tennant
  D~A, Chakoumakos B~C and Nagler S~E 2016 {\em Phys. Rev. B\/} {\bf 93}(13)
  134423

\bibitem{Gretarsson2013}
Gretarsson H, Clancy J~P, Liu X, Hill J~P, Bozin E, Singh Y, Manni S, Gegenwart
  P, Kim J, Said A~H, Casa D, Gog T, Upton M~H, Kim H~S, Yu J, Katukuri V~M,
  Hozoi L, van~den Brink J and Kim Y~J 2013 {\em Phys. Rev. Lett.\/} {\bf
  110}(7) 076402

\bibitem{Agrestini2017}
Agrestini S, Kuo C~Y, Ko K~T, Hu Z, Kasinathan D, Vasili H~B, Herrero-Martin J,
  Valvidares S~M, Pellegrin E, Jang L~Y, Henschel A, Schmidt M, Tanaka A and
  Tjeng L~H 2017 {\em Phys. Rev. B\/} {\bf 96}(16) 161107

\bibitem{Sears2015}
Sears J~A, Songvilay M, Plumb K~W, Clancy J~P, Qiu Y, Zhao Y, Parshall D and
  Kim Y~J 2015 {\em Phys. Rev. B\/} {\bf 91}(14) 144420

\bibitem{Dvorak2016}
Dvorak J, Jarrige I, Bisogni V, Coburn S and Leonhardt W 2016 {\em Review of
  Scientific Instruments\/} {\bf 87} 115109

\bibitem{Jarrige2018}
Jarrige I, Bisogni V, Zhu Y, Leonhardt W and Dvorak J 2018 {\em Synchrotron
  Radiation News\/} {\bf 31} 7--13

\bibitem{Chaloupka2016}
Chaloupka J and Khaliullin G 2016 {\em Phys. Rev. B\/} {\bf 94}(6) 064435

\bibitem{Sandilands2016}
Sandilands L~J, Tian Y, Reijnders A~A, Kim H~S, Plumb K~W, Kim Y~J, Kee H~Y and
  Burch K~S 2016 {\em Phys. Rev. B\/} {\bf 93}(7) 075144

\bibitem{Yadav2016}
Yadav R, Bogdanov N~A, Katukuri V~M, Nishimoto S, Van Den~Brink J and Hozoi L
  2016 {\em Scientific reports\/} {\bf 6} 37925

\bibitem{Johnson2015}
Johnson R~D, Williams S~C, Haghighirad A~A, Singleton J, Zapf V, Manuel P,
  Mazin I~I, Li Y, Jeschke H~O, Valent\'{\i} R and Coldea R 2015 {\em Phys.
  Rev. B\/} {\bf 92}(23) 235119

\bibitem{Sears-thesis}
Sears J~A 2017 {\em Neutron and X-ray Diffraction Studies of Structure and
  Magnetism in $\alpha$-RuCl$_3$\/} Ph.D. thesis University of Toronto

\bibitem{Jungho2012}
Kim J, Casa D, Upton M~H, Gog T, Kim Y~J, Mitchell J~F, van Veenendaal M,
  Daghofer M, van~den Brink J, Khaliullin G and Kim B~J 2012 {\em Phys. Rev.
  Lett.\/} {\bf 108}(17) 177003

\bibitem{Jungho2014}
{Kim} J, {Daghofer} M, {Said} A~H, {Gog} T, {van den Brink} J, {Khaliullin} G
  and {Kim} B~J 2014 {\em Nature Communications\/} {\bf 5} 4453

\bibitem{Borgwardt2019}
Borgwardt N 2019 {\em Optics on materials with strong spin-orbit coupling:
  topological insulators Bi2-xSbxTe3-ySey and the j=1/2 compounds Na2IrO3 and
  alpha-RuCl3\/} Ph.D. thesis Universit{\"a}t zu K{\"o}ln
  \urlprefix\url{https://kups.ub.uni-koeln.de/9187/}

\bibitem{Plotnikova2016}
Plotnikova E~M, Daghofer M, van~den Brink J and Wohlfeld K 2016 {\em Phys. Rev.
  Lett.\/} {\bf 116}(10) 106401

\bibitem{Porterfield2013}
Porterfield W 2013 {\em Inorganic Chemistry\/} (Elsevier Science) ISBN
  9780323138949

\bibitem{Reschke2017}
Reschke S, Mayr F, Wang Z, Do S~H, Choi K~Y and Loidl A 2017 {\em Phys. Rev.
  B\/} {\bf 96}(16) 165120

\bibitem{Kuiper1998}
Kuiper P, Guo J~H, S\aa{}the C, Duda L~C, Nordgren J, Pothuizen J~J~M, de~Groot
  F~M~F and Sawatzky G~A 1998 {\em Phys. Rev. Lett.\/} {\bf 80}(23) 5204--5207

\bibitem{Lai2014}
Lai C~H, Fung H~S, Wu W~B, Huang H~Y, Fu H~W, Lin S~W, Huang S~W, Chiu C~C,
  Wang D~J, Huang L~J, Tseng T~C, Chung S~C, Chen C~T and Huang D~J 2014 {\em
  Journal of Synchrotron Radiation\/} {\bf 21} 325--332

\bibitem{Brookes2018}
Brookes N, Yakhou-Harris F, Kummer K, Fondacaro A, Cezar J, Betto D, Velez-Fort
  E, Amorese A, Ghiringhelli G, Braicovich L, Barrett R, Berruyer G, Cianciosi
  F, Eybert L, Marion P, van~der Linden P and Zhang L 2018 {\em Nuclear
  Instruments and Methods in Physics Research Section A: Accelerators,
  Spectrometers, Detectors and Associated Equipment\/} {\bf 903} 175 -- 192
  ISSN 0168-9002

\bibitem{I21}
Diamond light source: I21 beamline homepage
  \url{https://www.diamond.ac.uk/Instruments/Magnetic-Materials/I21.html}

\bibitem{Veritas}
Max iv: Veritas beamline homepage
  \url{https://www.maxiv.lu.se/accelerators-beamlines/beamlines/veritas/}

\bibitem{Wray2015}
Wray L~A, Huang S~W, Jarrige I, Ikeuchi K, Ishii K, Li J, Qiu Z~Q, Hussain Z
  and Chuang Y~D 2015 {\em Frontiers in Physics\/} {\bf 3} 32 ISSN 2296-424X

\bibitem{Bambynek1972}
Bambynek W, Crasemann B, Fink R~W, Freund H~U, Mark H, Swift C~D, Price R~E and
  Rao P~V 1972 {\em Rev. Mod. Phys.\/} {\bf 44}(4) 716--813

\bibitem{Halasz2016}
Hal\'asz G~B, Perkins N~B and van~den Brink J 2016 {\em Phys. Rev. Lett.\/}
  {\bf 117}(12) 127203

\bibitem{Ament2011}
Ament L~J~P, van Veenendaal M, Devereaux T~P, Hill J~P and van~den Brink J 2011
  {\em Rev. Mod. Phys.\/} {\bf 83}(2) 705--767

\bibitem{Suzuki2019}
Suzuki H, Gretarsson H, Ishikawa H, Ueda K, Yang Z, Liu H, Kim H, Kukusta D,
  Yaresko A, Minola M, Sears J~A, Francoual S, Wille H~C, Nuss J, Takagi H, Kim
  B~J, Khaliullin G, Yavas H and Keimer B 2019 {\em Nature Materials\/} {\bf
  18} 563--567 ISSN 1476-4660

\bibitem{Gretarsson2019}
Gretarsson H, Suzuki H, Kim H, Ueda K, Krautloher M, Kim B~J,
  Yava\ifmmode~\mbox{\c{s}}\else \c{s}\fi{} H, Khaliullin G and Keimer B 2019
  {\em Phys. Rev. B\/} {\bf 100}(4) 045123

\end{thebibliography}
\end{document}